\documentclass[preprint,showpacs,amssymb,aps,pra]{revtex4}
\usepackage{graphicx}

\begin{document}

\title{
On the Inequivalence 
\\
of Renormalization and Self-Adjoint Extensions
\\ 
for Quantum Singular Interactions
}

\author{
 Horacio E. Camblong,$^{a}$\footnotetext{{\em E-mail
address\/}: camblong@usfca.edu (H.E. Camblong)}
Luis N. Epele,$^{b}$ Huner Fanchiotti,$^{b}$
 Carlos A. Garc\'{\i}a Canal,$^{b}$
and 
Carlos R. Ord\'{o}\~{n}ez$^{c,d}$}

\affiliation{
$^{a}$ Department of Physics, University of San Francisco, San
Francisco, California 94117-1080, USA \\
$^{b}$  Laboratorio de F\'{\i}sica Te\'{o}rica,
 Departamento de F\'{\i}sica,
Facultad de Ciencias Exactas,
Universidad Nacional de La Plata,
 C.C. 67--1900 La Plata, Argentina
\\
$^{c}$ Department of Physics, University of Houston, Houston,
Texas 77204-5506, USA
\\
$^{d}$
World Laboratory Center for Pan-American Collaboration in Science and
Technology,
\\
University of Houston Center, Houston, Texas 77204-5506, USA}

\begin{abstract}
 A unified S-matrix framework of quantum singular interactions is presented 
for the comparison of self-adjoint extensions and physical renormalization.
For the long-range conformal interaction
the two methods are not equivalent,
with renormalization acting as selector of a preferred extension
and regulator of the unbounded Hamiltonian.
\end{abstract}

\pacs{03.65.-w, 03.65.Ca, 11.10.Gh, 04.70.-s
\\
{\em Keywords:\/}  Renormalization; Self-adjoint extensions; 
Singular interactions; Conformal quantum mechanics; Quantum anomalies; 
S-matrix.}

\maketitle

\section{\large\bf I\lowercase{ntroduction}}
\label{sec:intro}

Singular quantum-mechanical systems involve interactions whose strongly 
divergent 
behavior at a given point governs the leading
physics~\cite{spector_RMP}.
This entails a hierarchy in which the usual 
preponderance of ``kinetic terms'' (the Laplacian and
concomitant centrifugal potentials)
is suppressed by the near-singularity dominance of the interaction.
In turn, the singular behavior implies an indeterminacy in the 
boundary condition at the singularity~\cite{spector_RMP,landau:77}.
Consequently, the standard technology of regular quantum mechanics
is supplemented by a mandatory regularization toolbox: 
either Von Neumann's method of self-adjoint 
extensions~\cite{Von-Neumann:29,self-adjoint_intro,reed-simon,albeverio},
 which addresses the boundary-value problem,
or
field-theory renormalization~\cite{jackiw-beg:91,huang:92,tarrach_gupta,camblong:isp_letter-dtI&II,renormalization_CQM,EFT_singular}, which deals with the underlying ultraviolet cause of such indeterminacy.

A crucial question in the theory of singular potentials
is whether these two apparently distinct methods are equivalent or not. 
In Ref.~\cite{jackiw-beg:91},
their equivalence was shown for the delta-function conformal interaction.
However, the more complex issue of their equivalence for the long-range 
conformal interaction has not been exhaustively studied.
Moreover, in the absence of a systematic comparison,
it is usually conjectured that both methods have comparable efficacy and 
ultimately yield solutions in one-to-one correspondence.

In this paper we show that the above ``equivalence conjecture''
is incorrect, within the scope of traditional physical regularizations,
for which the singularity emerges
from a $(0+1)$-dimensional {\em effective\/} field theory~\cite{EFT}.
Specifically, the generalization of Ref.~\cite{jackiw-beg:91} breaks down due
to the spectral properties of long-range singular interactions.
We analyze these issues 
within a comparative S-matrix framework of 
self-adjoint-extensions and renormalization.
Furthermore, for the conformal interaction, we show that: 
(i) physical regularization selects a preferred self-adjoint extension
for ``medium-weak coupling;''
(ii) the unbounded nature of the Hamiltonian  for ``strong coupling''
can only be fixed by physical renormalization---otherwise, a spectrum
not bounded from below would yield an unstable system. 

\section{\large\bf S\lowercase{-matrix framework 
for singular quantum mechanics}}
\label{sec:S-matrix_Singular-QM}

For the complete characterization of the physics of a singular 
system, it proves useful to introduce a unified S-matrix framework.
In addition to generating its built-in scattering observables, 
the S-matrix permits a comprehensive analysis
of the spectrum of singular potentials, 
with the bound-state sector displayed through its poles.
Specifically, for the family of {\em long-range singular interactions\/}
$V({\bf r}) = - \lambda
\,
 r^{-\gamma}$ (with $\gamma \geq 2$),
the multidimensional wave function 
$
\Psi ({\bf r})
=  
Y_{l m}( {\bf \Omega})
u(r)/r^{(d-1)/2} $
in $d = 2 (\nu + 1) $ spatial dimensions leads to
\begin{equation}
\left[
\frac{d^{2}}{dr^{2}} 
\,
+ k^{2} 
+
\frac{ \lambda
}{r^{\gamma}}
-
\frac{ \left( l + \nu \right)^{2} - 1/4
}{r^{2}}
\right]  
u (r) = 0
\;  
\label{eq:singular-QM_radial}
\end{equation}
(in natural units~\cite{camblong:isp_letter-dtI&II}). 
The observables are encoded in the S-matrix through 
the asymptotics of two independent solutions
$u_{1,2}(r)$
of Eq.~(\ref{eq:singular-QM_radial}),
 with
\begin{equation}
u (r) 
\propto 
\hat{S} 
\, 
u_{1} (r) +  u_{2} (r) 
\; ,
\label{eq:wave-function_asymptotic-basis}
\end{equation}
where we will adopt the convention
\begin{equation}
u_{1,2} (r)
\stackrel{( r \rightarrow \infty )}{\propto}
\frac{1}{ \sqrt{k} }
\,
e^{\pm i kr} e^{\mp i \pi/4}
\label{eq:asymptotic-wave-normalization}
\end{equation}
(up to a real numerical factor)
leading to the factorization
\begin{equation}
S = e^{i \pi \left( l+\nu \right)} \, \hat{S} 
\; 
\label{eq:S-matrix_and_reduced-S-matrix}
\end{equation}
for the usual definition of S-matrix.

In addition, two convenient 
solutions $u_{\pm}(r)$ 
of Eq.~(\ref{eq:singular-QM_radial})
can be introduced as ``singularity probes,'' 
to fully capture the characteristic behavior of the theory as $r \sim 0$.
They can be defined in terms of the leading WKB behavior
near the origin~\cite{landau:77},
which is ``asymptotically exact''~\cite{semiclassical_BH}.
Correspondingly,
\begin{equation}
u  (r) \propto 
\Omega 
\, u_{+}(r) +  u_{-} (r)
\; ,
\label{eq:wave-function_near-origin-basis}
\end{equation}
where $\Omega $ is a ``singularity parameter.''
At the practical level,
Eqs.~(\ref{eq:wave-function_asymptotic-basis})
and
(\ref{eq:wave-function_near-origin-basis})
represent two different resolutions
of the wave function~\cite{esposito},
which can be compared, provided that the basis expansions
\begin{equation}
u_{j} (r)= \alpha_{j}^{\sigma} 
\,
u_{\sigma}(r)
\label{eq:basis-transformation}
\end{equation}
(where the summation convention is adopted,
with $j=1,2$, and $\sigma = \pm$)
are established.
Thus, the relation between the ``components'' 
$\hat{S}$ in Eq.~(\ref{eq:wave-function_asymptotic-basis})
and
$\Omega$ in Eq.~(\ref{eq:wave-function_near-origin-basis})
follows by inversion of the transfer matrix
$\left[ \alpha_{j}^{\sigma} \right]$.
In particular, the reduced S-matrix  $\hat{S} = \hat{S} (\Omega)$ 
is given by the fractional linear transformation
\begin{equation}
\hat{S}
=
\frac{ \alpha_{2}^{-} \,
\Omega -  \alpha_{2}^{+}  }{
-  \alpha_{1}^{-} \, \Omega +  \alpha_{1}^{+}  }
\; ,
\label{eq:S-matrix_singular}
\end{equation}
which will play a crucial role for the remainder of the paper.

\section{\large\bf S\lowercase{elf-adjoint extensions:}
 C\lowercase{onformal}
 S\lowercase{-matrix}}
\label{sec:CQM_self-adjoint}

The main goal of our paper is to highlight the failure of the 
``equivalence conjecture'' for long-range singular interactions.
As we will see, this in part due to the fact that
the extensions do not describe a unique physical system but 
an {\em ensemble\/} of systems labeled by extension parameters.
Thus, the selection of the relevant solution involves identifying 
the appropriate physical system within the ensemble; in short, 
as stated in Ref.~\cite{reed-simon}, this is not a mathematical 
``technicality'' but is to be constrained by the physics.

Let us start by stating a singular potential 
problem~\cite{spector_RMP} within the method of 
self-adjoint extensions~\cite{self-adjoint_intro}. For the case of 
central symmetry, this reduces to Eq.~(\ref{eq:singular-QM_radial}),
whose solutions may
involve an ensemble of Hamiltonians rather than a 
single-system Hamiltonian, thereby leading to the 
physical indeterminacy described above.
In this section, we will demonstrate the nature of this 
problem by solving the Schr\"{o}dinger
Eq.~(\ref{eq:singular-QM_radial}) for $\gamma =2 $,
i.e., for the long-range conformal interaction
$V({\bf r}) = -\lambda/r^{2}$.
In our approach we will subsume the results of Ref.~\cite{narnhofer:74}
within an effective-field theory interpretation and we will give 
further support to our conclusions using established physical applications.
Incidentally, with an appropriate interpretation,
the quantum-mechanical conformal analysis 
{\em transcends nonrelativistic quantum mechanics\/} in an effective reduced 
form that includes applications to the near-horizon physics 
and thermodynamics of black holes with generalized Schwarzschild 
metrics~\cite{near_horizon,BH_thermo_CQM,semiclassical_BH,padmanabhan}, 
and also in gauge theories~\cite{gusynin1,renormalization_CQM}.

In our analysis, we will make use of two distinctive features
of the conformal interaction: 
(i) the existence of a critical coupling;
(ii) its SO(2,1) conformal 
symmetry~\cite{alfaro_fubini_furlan:76,renormalization_CQM,jackiw-ISP:72}.
The  critical coupling $\lambda^{(*)}= (l + \nu)^{2}$, which separates 
two distinct coupling regimes~\cite{camblong:isp_letter-dtI&II},
is associated with a qualitative change of the solutions of 
Eq.~(\ref{eq:singular-QM_radial}) and its multidimensional counterpart 
for $\gamma =2$. In effect, the radial wave functions
$
  u (r) 
=
r^{1/2}
\,
Z_{s} ( k \, r )
$
involve a Bessel function $Z_{s}(z) $ of order $s$, whose 
nature changes abruptly when the parameter
\begin{equation}
s^{2} =
\lambda^{(*)} - \lambda 
\label{eq:def_Bessel-index}
\end{equation}
goes through zero.
In this paper we show that there are {\em no additional physical regimes\/}
under reasonable conditions supported by a large class of realizations. 
However, as we will see next, 
this simple description is altered by
the method of self-adjoint extensions, which generically gives
rise to an additional transitional coupling regime.
 Specifically, the physical indeterminacy of Von Neumann's method takes 
central stage for the conformal Hamiltonian with $0 < s < 1$, which 
admits a one-parameter family of self-adjoint extensions~\cite{narnhofer:74}, 
with
\begin{equation}
\lambda^{(*)}  -1
< 
\lambda
<
\lambda^{(*)}
\; 
\label{eq:medium-weak_coupling}
\end{equation}
defining a subcritical {\em medium-weak coupling\/}.
The existence of this additional regime for self-adjoint extensions
 can be easily seen by applying the standard 
technique~\cite{Von-Neumann:29,self-adjoint_intro,reed-simon}, 
 according to which the extensions are given through the eigenfunctions 
$u_{\pm}^{({\rm SA})} (r) $, 
 with eigenvalues $\pm \mu^{2}i$, where $\mu$
is an arbitrary scale~\cite{floating_point_scale}. 
The corresponding deficiency subspaces, with dimensionalities 
$d_{\pm}$, extend the operator domain to guarantee self-adjointness, 
when $d_{+}=d_{-}$.

In what follows, and for subsequent comparison with the renormalized
solutions, we will use the superscript notation in 
$u_{\pm}^{({\rm SA})} (r) $ 
(and similar expressions for other quantities)
to refer to entities computed by means of self-adjoint extensions.
For the conformal medium-weak coupling~(\ref{eq:medium-weak_coupling}),
the asymptotics (for $ r \sim \infty$) yields the formal solutions
\begin{equation}
u_{\pm}^{({\rm SA})} (r) 
=
\sqrt{r}
 H^{(1,2)}_{s} (e^{\pm i \pi/4} \mu r)
\; ,
\end{equation}
in terms of Hankel functions,
each with multiplicity one, so that $d_{+}=d_{-}=1 $.
Therefore, for each $z \in [0,2 \pi)$, the extension
\begin{equation}
u^{({\rm SA})} (r) 
=
u_{+}^{({\rm SA})} (r) 
+ 
e^{iz}
u_{-}^{({\rm SA})} (r) 
\; 
\label{eq:wave-function_SA-extension-basis}
\end{equation}
defines a specific boundary condition
by comparison with the conformal building blocks 
$ u_{\pm} (r) $ near the singularity,
through Eq.~(\ref{eq:wave-function_near-origin-basis}).
In this section, the appropriate singularity parameter will be denoted
by $\Omega^{({\rm SA})} (z)$.
Explicitly, this comparison entails
\begin{equation}
u (r) 
\propto
\Omega^{({\rm SA})} (z)
\, 
 u_{+} (r) 
+ 
u_{-} (r) 
\stackrel{( r \rightarrow 0 )}{\propto}
u^{({\rm SA})} (r) 
\; ,
\label{eq:wave-function_SA-extension-basis-to-near-sing}
\end{equation}
where 
$u_{\pm}(r) \propto \sqrt{r} \, J_{\pm s} (kr)$
are given in terms of the ordinary Bessel functions.
The identification 
in Eq.~(\ref{eq:wave-function_SA-extension-basis-to-near-sing})
is made to leading order as the singularity is approached;
specifically, from
the small-argument expansions of Bessel and Hankel functions, 
\begin{equation}
u^{({\rm SA})} (r) 
\stackrel{( r \rightarrow 0 )}{\propto}
\frac{ 2 \, e^{i z/2} }{ \sin \left( \pi s \right) }
\,
\sqrt{r}
\left[
 F_{+}^{({\rm SA})} (z)
\,
 \left( 
\frac{\mu}{k}
 \right)^{s}
\,
J_{s} (k r)
-
F_{-}^{({\rm SA})} (z)
\,
\left( 
\frac{\mu}{k}
\right)^{-s}
\,
J_{-s} (k r)
\right]
\; ,
\label{eq:VonNeumann-building-block-expansion}
\end{equation}
where
\begin{equation}
F_{\pm}^{({\rm SA})} (z) =
 \sin 
\left( 
\frac{z}{2} +  
\phi_{\pm} (s)
\right)
\; 
\label{eq:F-plus-minus}
\end{equation}
and
$\phi_{\pm} (s) 
=
\pi s/2  \pm  \pi s/4 
$.
Comparison with Eq.~(\ref{eq:wave-function_SA-extension-basis-to-near-sing})
shows that
\begin{equation}
\Omega^{({\rm SA})} (z)
= 
-
\varrho^{({\rm SA})} (z) 
\, 
\left( 
\frac{k}{\mu }
 \right)^{-2s} 
\; ,
\label{eq:Omega-ISP_SA}
\end{equation}
where
\begin{equation}
\varrho^{({\rm SA})} (z) 
\! = \! 
\frac{
F_{+}^{({\rm SA})} (z) }{
F_{-}^{({\rm SA})} (z) }
\; .
\label{eq:S-matrix_aux-ratio_SA}
\end{equation}

Furthermore, the observables can be retrieved from the S-matrix,
whose form can be derived through the solutions $u_{1,2}(r)$,
Eq.~(\ref{eq:wave-function_asymptotic-basis}).
With the normalization~(\ref{eq:asymptotic-wave-normalization}),
the corresponding conformal building blocks are
\begin{equation}
u_{1,2}(r)
= e^{\pm is\pi/2} \, \sqrt{\frac{\pi \, r }{ 2} }  \, H^{(1,2)}_{s } (kr)
\; .
\end{equation}
In a similar manner, the conformal singularity probes
\begin{equation}
u_{\pm}(r) = \sqrt{ \frac{ i \pi r }{ \sin \left( \pi s \right)} } 
\, J_{\pm s} (kr)
\; ,
\label{eq:CQM_singularity-probes}
\end{equation}
needed in Eq.~(\ref{eq:wave-function_near-origin-basis}), 
provide the transformation coefficients 
for Eq.~(\ref{eq:basis-transformation}),
\begin{equation}
 \alpha_{1}^{\pm}
=
  \alpha_{2}^{\mp}
=
 \pm  \sqrt{ \frac{ i  }{ 2 \sin (\pi s) } }
\,
e^{\mp i \pi s/2}
\; .
\label{eq:transf-matrix_CQM}
\end{equation}
Due to conformal invariance, 
Eqs. (\ref{eq:CQM_singularity-probes}) and (\ref{eq:transf-matrix_CQM})
are obviously defined up to normalization factors, 
as will be discussed elsewhere.
Then, the S-matrix---derived from Eqs.~(\ref{eq:S-matrix_singular}) and 
(\ref{eq:transf-matrix_CQM})---takes the generic conformal expression
\begin{equation}
\tilde{S}
\equiv
\hat{S}
\, e^{i \pi s}
=
\frac{  e^{i \pi s} + \Omega  }{
 e^{- i \pi s} + \Omega  }
\; ,
\label{eq:S-matrix_singular-conformal}
\end{equation}
which will also be used in the next section.
Moreover, in Eq.~(\ref{eq:S-matrix_singular-conformal}),
the required singularity parameter
$\Omega
\equiv 
\Omega^{({\rm SA})} (z) 
$
is given in Eq.~(\ref{eq:Omega-ISP_SA}).
As a result, the {\em self-adjoint conformal S-matrix \/} becomes
\begin{equation}
\tilde{S}^{({\rm SA})}  
\equiv
\hat{S}^{({\rm SA})}  
\, e^{i \pi s}
=
\frac{ e^{i \pi s} 
\, 
\left( k/\mu \right)^{2s} 
- \varrho^{({\rm SA})} (z)  }{
e^{-i \pi s} \, \left( k / \mu   \right)^{2s} 
- \varrho^{({\rm SA})} (z)  }
\; .
\label{eq:S-matrix_SA}
\end{equation}
Furthermore, in addition 
to the familiar arbitrariness in $\mu$~\cite{floating_point_scale},
Eq.~(\ref{eq:S-matrix_SA}) 
displays the free parameter $z$,
which poses a physical indeterminacy.

\begin{figure}
\centering

\resizebox{4.25in}{!}{\includegraphics{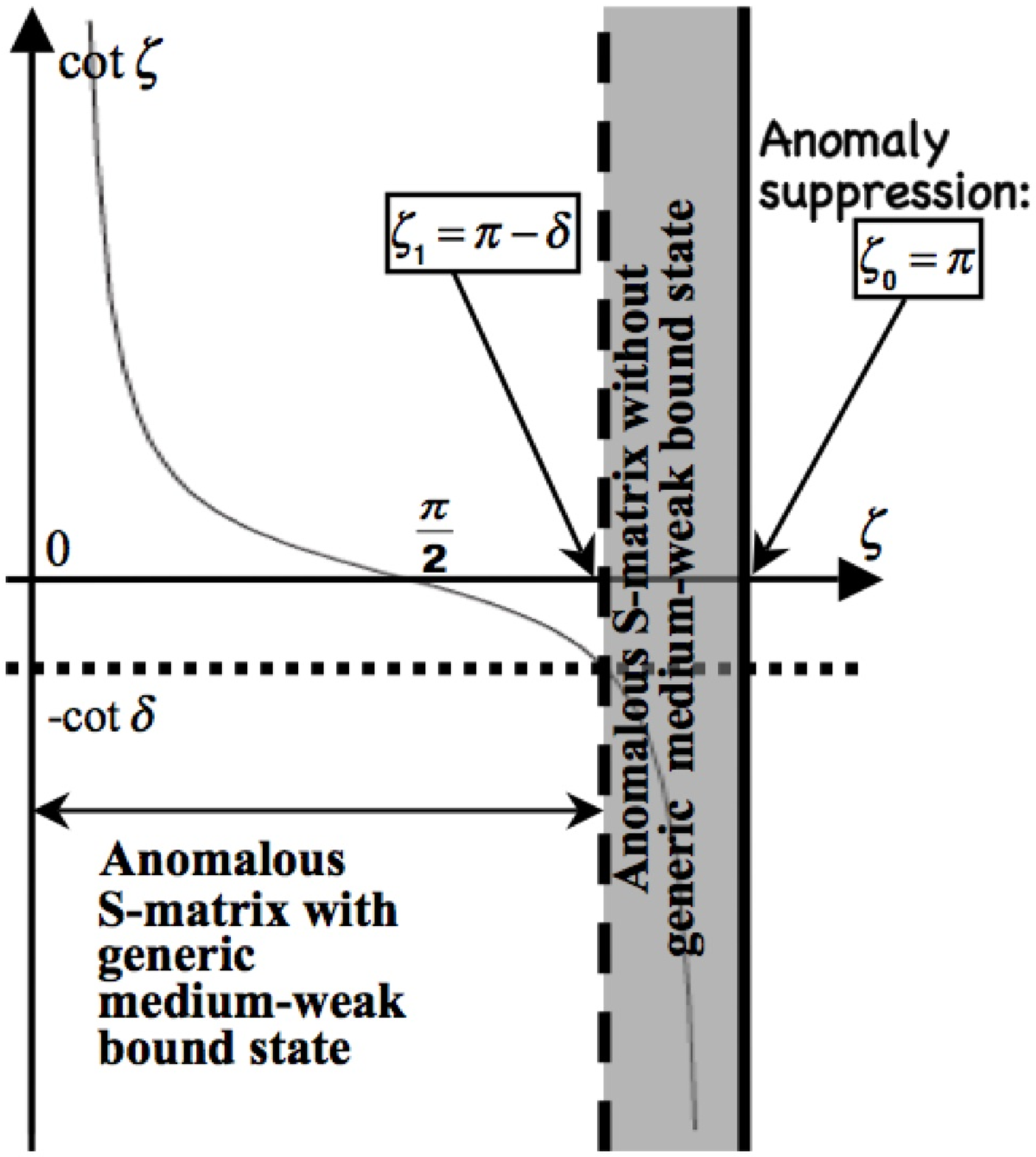}}

\caption{Self-adjoint extensions in the medium-weak coupling regime.
The parameters 
$\delta \equiv \pi s/2$ (with $0 < \delta < \pi/2$)
and
$\zeta \equiv \left( z + \delta \right)/2 \pmod{ 2 \pi} $
are defined. 
The region ``without generic medium-weak bound state''
corresponds to  
$\varrho^{({\rm SA})} (z) 
=
F_{+}^{({\rm SA})} (z)/ F_{-}^{({\rm SA})} (z) 
< 0 $
[see Eqs.~(\ref{eq:F-plus-minus}) and
(\ref{eq:S-matrix_aux-ratio_SA})];
strictly speaking, for some rational values of $s$,
this region may admit a 
bound state as well.
The value $z=z_{0}$ effectively suppresses the
medium-weak anomaly.}

\label{fig:SA_cotangent-graph1}

\end{figure}

In a similar manner, for the bound-state sector,
the normalizable function
$
u (r) \! \propto \! \sqrt{r} K_{s} (\kappa r)
$
is a possible candidate for a 
 ``medium-weak bound state,'' with energy
\begin{equation}
E_{0} = 
- \mu^{2}
\left[
{\varrho}^{({\rm SA})} (z) 
\right]^{1/s}
=
- \mu^{2}
\left[ \frac{ \sin \left( z/2 + 3 \pi s/4 \right) }{
\sin \left( z/2 +  \pi s/4 \right) } \right]^{1/s}
\;  ,
\label{eq:SA_energies_intermediate-weak}
\end{equation}
which can be obtained by comparison with
the extensions~(\ref{eq:wave-function_SA-extension-basis})
near the singularity.
These states,
as shown in Fig.~\ref{fig:SA_cotangent-graph1},
 exist for
$ 
z 
\in 
{\mathcal S} 
\equiv 
[0,z_{1}) \cup  (z_{0}, 2 \pi)
$,
independently of the value of $0< s < 1$,
where $z_{1} = 2 \pi \! - \! 3\pi \, s/2$
and $z_{0}= 2 \pi \! - \! \pi \, s/2$
are the formal limits of 
$\kappa =0$ and $\kappa =\infty$ respectively.
Curiously, their existence is maintained 
even in the cases when the potential is repulsive.

A more careful analysis of the S-matrix~(\ref{eq:S-matrix_SA}) 
reveals additional properties of the self-adjoint-extended Hamiltonians.
The poles of Eq.~(\ref{eq:S-matrix_SA})
in the complex $k$ plane are located at the points
\begin{equation}
k_{n} = 
i
\mu 
\left[
{\varrho}^{({\rm SA})} (z) 
\right]^{1/2s}
\, e^{n \pi i/s}  
\; ,
\label{eq:S-matrix_poles}
\end{equation}
where $n$ is an integer, and
the sign of 
${\varrho}^{({\rm SA})} (z) $ in 
$k_{0} 
= 
i \mu  \left[ {\varrho}^{({\rm SA})} (z) \right]^{1/2s} $
 shows again the separation of the interval $z \in [0,2 \pi)$ in two parts:
$\mathcal S$, where it is positive,
 and its complement $ {\mathcal T} \equiv (z_{1},z_{0}) $, 
where it is negative. The first region, $\mathcal S$, is seen to 
reproduce the bound state energy~(\ref{eq:SA_energies_intermediate-weak})
from $E_{0} = k_{0}^{2} $, with $n=0$ in Eq.~(\ref{eq:S-matrix_poles}).
However, the multiplicity factor $e^{2 i n \delta}$ in
Eq.~(\ref{eq:S-matrix_poles}), with $\delta = \pi s/ 2 $,
typically generates additional poles; these are located
symmetrically at points of the same modulus with a phase
increment $2 \delta$.
Whenever $ 2 \delta$ is a rational number times $ 2 \pi $, 
which occurs for rational values of $s$, these form
a finite set; however, for irrational values of $s$,
they consist of an infinite set of distinct poles.
The second region, $\mathcal T$, involves a principal
value $k_{0}= 
i \mu   \left| {\varrho}^{({\rm SA})} (z) \right|^{1/2s} \,
 e^{i \delta}$, which typically
lies in the complex plane and away from the positive imaginary axis;
in addition, the multiplicity factor 
$e^{2 i n \delta}$ unfolds a similar pole structure as for 
$\mathcal S$, but rotated through a phase $\delta$.
Most of these poles, for $z$ values both in
$\mathcal S$ and in $\mathcal T$,
do not have a simple interpretation; however, some of them 
may be interpreted as resonances, while
occasionally, for specific rational values of $s$, some may generate 
a virtual state or even a single bound state in $\mathcal T$
[with the latter occurring for
$s = (2p-1)/4 q$, when $p$ and $q$ are positive integers,
such that $s<1$].
It should be noticed that this polology of the self-adjoint 
extensions exists for all values of $z$ except for $z_{0}$ and $z_{1}$.

\section{\large\bf 
E\lowercase{ffective-field-theory approach:}
C\lowercase{onformal}
S\lowercase{-matrix}}
 \label{sec:CQM_S-matrix}

In this section we will address the main theme of our paper:
the resolution of the physical indeterminacy of self-adjoint extensions
via physical regularization and renormalization.
In simple terms, while the self-adjoint method deals with the 
symptoms of the ``quantum illness'' at the level of the 
boundary conditions, only the effective field approach 
tackles the underlying cause of the singular behavior. 

Furthermore, the S-matrix framework of Sec.~\ref{sec:S-matrix_Singular-QM}
serves as the analytical basis for a comprehensive study of singular potentials
and will be central to our proof of the inequivalence of the two methods for 
the long-range conformal interaction. The basic procedure was carried out
 in the previous section using self-adjoint extensions. 
In this section, following a similar methodology,
we will derive the more stringent conditions 
arising from a regularization and renormalization of the conformal potential.

\subsection{\large\it
Regularized conformal S-matrix and coupling regimes}

As anticipated in Sec.~\ref{sec:intro},
the presence of a singularity requires an effective treatment. 
In this renormalization context, singular quantum mechanics calls for 
the use of a cutoff $a$ and
an ultraviolet regularizing core $ V^{(<)}({\bf r}) $ for $ r \alt a$.
As a result, the functions $u_{\pm}(r)$ can be used to
provide the required matching through the dimensionless logarithmic 
derivative~\cite{renormalization_CQM}
\begin{equation}
{\mathcal L} 
\equiv
\left.
r \,
\frac{ d  \ln \left[ u(r)/ r^{\gamma/4}  \right]
 }{dr}  
\right|_{r=a}
\; .
\label{eq:log-derivative}
\end{equation}
In essence, Eq.~(\ref{eq:log-derivative}) effects the regularization, as 
shown below for conformal quantum mechanics and other singular interactions. 
 It should be noticed that the unconventional normalization 
of Eq.~(\ref{eq:log-derivative}) is specifically tailored to the 
computations for the conformal and other singular interactions.

In this paper we show that there are no additional physical regimes,
other than the weak and strong,
under reasonable conditions supported by a large class of realizations. 
In effect, from the regularization procedure,
the coefficient $\Omega $ can be computed as a function of $a$,
\begin{equation}
\Omega (a) 
= 
-
\rho 
\, 
\left( 
\frac{k a }{2}
 \right)^{-2s} 
\; ,
\label{eq:Omega-ISP_regularized}
\end{equation}
with 
\begin{equation}
\rho = \frac{F_{+} }{ F_{-} }
\; 
\label{eq:S-matrix_aux-ratio_CQM}
\end{equation}
given in terms of
\begin{equation}
F_{\pm} =
\Gamma ( 1 \pm  s) 
\;
\left(
\frac{ {\mathcal L} }{ s } \pm 1
\right)
\; .
\label{eq:S-matrix_rho-ratio-coeff}
\end{equation}
From the generic Eq.~(\ref{eq:S-matrix_singular-conformal}),
the singularity parameter~(\ref{eq:Omega-ISP_regularized}) yields the 
{\em regularized S-matrix\/}
\begin{equation}
\tilde{S}
\equiv
\hat{S}
\, e^{i \pi s}
=
\frac{ e^{i \pi s} 
\, 
\left( 
k a/2
 \right)^{2s} - \rho  }{
e^{-i \pi s} 
\, 
\left( 
k a/2 
\right)^{2s} - \rho  }
\; .
\label{eq:S-matrix-CQM_regularized}
\end{equation}
Notice the formal similarity between Eqs.~(\ref{eq:S-matrix_SA}) and
(\ref{eq:S-matrix-CQM_regularized}),
Eqs.~(\ref{eq:Omega-ISP_SA}) and (\ref{eq:Omega-ISP_regularized}),
and between Eqs.~(\ref{eq:S-matrix_aux-ratio_SA}) and
(\ref{eq:S-matrix_aux-ratio_CQM}).
However, with the physical regulator $a$,
the functional form of 
Eq.~(\ref{eq:S-matrix-CQM_regularized}) shows the existence of 
a critical condition at the index value $s=0$, in the transition from 
positive real $s$ to an imaginary value, according to
Eq.~(\ref{eq:def_Bessel-index}).
Consequently, when $a \rightarrow 0$, 
{\em two and only two distinct regimes emerge\/}:
\\
(i) {\sf Weak or subcritical} 
($s$ real), 
with a  ``quasi-free S-matrix''
\begin{equation}
\left.
\tilde{S}
\right|_{\rm quasi \; free}
= 
\hat{S} e^{i \pi s} = 1
\; , 
\label{eq:quasi-free_S-matrix}
\end{equation}
for which the limit is well defined. 
By comparison with the
original generic solution, 
it is observed that 
the more divergent power, $u_{-}(r)$, 
in Eq.~(\ref{eq:wave-function_near-origin-basis})
is rejected and no 
outstanding indeterminacy remains.
\\
(ii) {\sf Strong or supercritical\/}
($s= i \Theta$ pure imaginary),
for which the regularized S-matrix~(\ref{eq:S-matrix-CQM_regularized}) 
has an ill-defined limit.
In contradistinction to the subcritical case, it is observed that the 
boundary-condition indeterminacy
cannot be resolved by the ordinary techniques of regular quantum mechanics.

\subsection{\large\it
Effective-field-theory resolution
of the physical indeterminacy of self-adjoint extensions}
\label{sec:renormalization-CQM_resolution}

The main conclusion of Sec.~\ref{sec:CQM_self-adjoint}
is the indeterminacy of the self-adjoint extension method, 
with the appearance of an unusual bound state for medium-weak coupling.
We will now show that this non-uniqueness can be physically resolved 
by adopting the effective approach outlined above. 

The selection of a physically preferred
self-adjoint extension can be established by comparing the 
coefficients in the expansions of the exterior solutions $u_{1,2}(r)$ for 
the self-adjoint extensions with the effective regularized solutions. 
This can be done step by step, verifying self-consistency, 
or more simply by direct comparison of the S-matrices for both approaches.
The latter procedure involves identifying Eq.~(\ref{eq:S-matrix_SA}) with 
Eq.~(\ref{eq:S-matrix-CQM_regularized}) in the medium-weak coupling regime,
with the effective S-matrix collapsing to 
the quasi-free form~(\ref{eq:quasi-free_S-matrix}).
Consequently, in Eq.~(\ref{eq:S-matrix_SA}) the choice
\begin{equation}
\left.
\tilde{S}
\right|_{\rm quasi \; free}
= 1
\Rightarrow 
 {\varrho}^{({\rm SA})} (z) 
 = \infty 
\Rightarrow 
 z_{0}= 2 \pi \! - \! \pi \, s/2
\; ,
\label{eq:correct_SA-extension}
\end{equation}
is enforced,
and a unique correct extension is selected by the effective
renormalization interpretation. 
The same conclusion is reached by direct comparison
of Eqs.~(\ref{eq:Omega-ISP_SA}) and (\ref{eq:Omega-ISP_regularized}).
In a simpler framework, this 
amounts to a straightforward regularization~\cite{meetz:64}
via the steps leading to Eq.~(\ref{eq:S-matrix-CQM_regularized}).
Moreover,
Eq.~(\ref{eq:VonNeumann-building-block-expansion})
shows that this is equivalent to the rejection of the 
more singular solution.
Thus, this procedure lifts the 
indeterminacy in the boundary condition;
in short, the extension with $z=z_{0}$ is selected 
by the physical regularization procedure.
At a purely formal level,
in terms of the bound-state sector, 
Eq.~(\ref{eq:correct_SA-extension}) corresponds to the
bound-state scale $\kappa \rightarrow \infty$;
no such bound state exists, but this procedure
gives the correct asymptotics,
$\Omega = - e^{- i \pi s}$ for a formal bound state
[with $S=\infty$ in Eq.~(\ref{eq:S-matrix_singular-conformal})]---such that 
$a \rightarrow 0$ and $\kappa \rightarrow \infty$  simultaneously in
$\Omega \propto \left( \kappa a/2 \right)^{-2s}$.

Incidentally, the SO(2,1) symmetry seems to play a 
selective role in this problem as well.
In effect, the self-adjoint-extended S-matrix~(\ref{eq:S-matrix_SA})
breaks the conformal symmetry for all self-adjoint extensions except
$z=z_{0}$ and $z=z_{1}$, as can be seen by inspection of
its explicit dependence with respect to the scale $k$.
The corresponding symmetry-breaking solutions can only be regarded unphysical
when this occurs in the presence of the symmetry-preserving extensions.
Moreover, the extension $z=z_{1}$ can be further excluded as it does not 
reproduce the correct S-matrix for the free particle in three dimensions.
This line of argumentation again shows that the correct physical
self-adjoint extension is provided by $z= z_{0}$.

In conclusion, the choice $z=z_{0}$ circumvents 
the {\em apparent quantum anomaly\/} of the S-matrix~(\ref{eq:S-matrix_SA}),
including the removal of a medium-weak bound state~\cite{Friedrichs}.
The symmetry breaking can be traced to
extraneous delta-function potentials~\cite{zorbas:80}, 
which are ordinarily not warranted by the long-range conformal interaction.
In this {\em medium-weak anomaly suppression\/}, when the phenomenology 
does not support extra interactions, the leading physics is governed by
symmetry and regularization. By contrast,
we will see that an anomaly cannot be circumvented for strong coupling.

\subsection{\large\it
The Nontrivial Physics of the Conformal Strong Coupling}

 We will now examine the physics of the effective renormalization 
procedure for strong coupling. As it turns out, 
their deceptively similar spectra hide profound physical differences, 
which are related to the precise origin of the quantum anomaly.

\begin{figure}
\centering

\resizebox{5.5in}{!}{\includegraphics{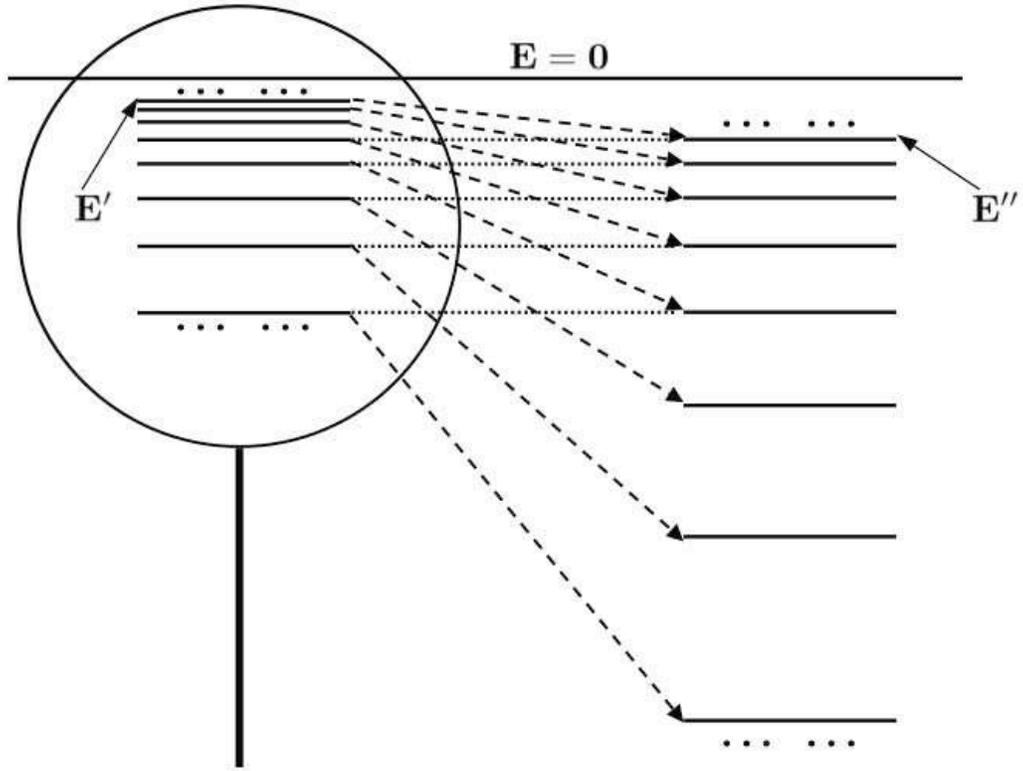}}

\caption{The strong-coupling conformal spectrum displays 
a residual discrete scale invariance, under magnifications $E''/E'$.
This is limited by {\em infrared and ultraviolet cutoffs.\/} }

\label{fig:conformal_spectrum}

\end{figure}

 Our comparative analysis is again displayed by the S-matrix.
In the supercritical case, 
with $s = i \Theta$, the function $\rho$ of the renormalization approach
in Eq.~(\ref{eq:S-matrix_aux-ratio_CQM})
turns into a pure phase because $F_{+}^{*} = - F_{-}$, as shown by 
Eq.~(\ref{eq:S-matrix_rho-ratio-coeff}).
Thus, Eq.~(\ref{eq:S-matrix-CQM_regularized}) gets replaced by
\begin{equation}
\tilde{S}
\equiv
\hat{S}
\, e^{- \pi \Theta}
=
\frac{ e^{- \pi \Theta} \, \left( k/\mu  \right)^{2 i  \Theta} 
+ e^{- 2 i \left( \Theta \, \gamma_{\Theta} +  \chi \right) } \,
\left( 2/\mu a \right)^{2 i  \Theta} 
}{
e^{ \pi \Theta} \, \left( k / \mu   \right)^{2 i \Theta} 
+
e^{ - 2 i 
\left( \Theta \, \gamma_{\Theta} +  \chi \right) } \,
\left( 2/\mu a \right)^{2 i  \Theta} 
}
\; ,
\label{eq:S-matrix_strong-regularized}
\end{equation}
where $\mu$ 
is an arbitrary scale,
$\gamma_{\Theta} = - 
\left\{
{\rm phase} 
\left[
\Gamma (1 +i\Theta )
\right]
\right\}/\Theta
$,
and 
$\chi = \pi \, f_{0} = 
{\rm phase} 
\left(
1 + i 
{\mathcal L}/\Theta 
\right)
$.
Then, Eq.~(\ref{eq:S-matrix_strong-regularized})
can be physically interpreted
by applying the well-known limit-cycle behavior that arises 
from the renormalization of the conformal potential
with a regularizing core~\cite{beane:01,braaten:04,hammer:06}
as $a \rightarrow 0$. 
This limit cycle is easily seen from the oscillatory nature 
of the imaginary powers in Eq.~(\ref{eq:S-matrix_strong-regularized}), 
with the core renormalization encoded in $\chi$
through the logarithmic derivative 
${\mathcal L}$, Eq.~(\ref{eq:log-derivative}).
 In this novel approach,
it is easy to see the emergence of nontrivial bound states 
through the poles of the S-matrix,
with a reference 
level $E_{0} = - \mu^{2}$ defined
in terms of the renormalization scale $\mu$.
Consequently,
Eq.~(\ref{eq:S-matrix_strong-regularized})
turns into the
{\em renormalized S-matrix\/}
\begin{equation}
\tilde{S}
\equiv
\hat{S}
\, e^{- \pi \Theta}
=
\frac{ e^{- \pi \Theta} \, 
 \left( k/\sqrt{|E_{0}| } \right)^{2 i  \Theta} 
- 1
}{
e^{ \pi \Theta} \, 
 \left( k/\sqrt{|E_{0}| } \right)^{2 i  \Theta} 
- 1
}
\; ,
\label{eq:S-matrix_renormalized}
\end{equation}
where 
$E_{0} = - (2 e^{- \gamma_{\Theta}}/a )^{2}
e^{ -2 \pi f_{0}/\Theta}
$.
In particular, from its  poles, the S-matrix
yields the bound state spectrum by considering the multiplicity 
arising from the imaginary nature of the exponent $i \Theta$ 
in Eq.~(\ref{eq:S-matrix_renormalized}). The ensuing energy levels are
\begin{equation}
E_{ n }
=
E_{0}
\,
\exp \left( - \frac{2 \pi n}{\Theta } \right)
\;  ,
\label{eq:energy-spectrum_CQM}
\end{equation}
whose 
most important property is {\em geometric scaling\/}, which is
{\em uniquely determined by discrete scale symmetry\/}:
Given two bound-state energies $E'$ and $E''$,
the spectrum is invariant under the magnification $E''/E'$
(see Fig.~\ref{fig:conformal_spectrum}).
Moreover, from Eq.~(\ref{eq:wave-function_asymptotic-basis}),
the corresponding bound states can be found by enforcing 
the condition $\hat{S} = \infty$,
which leads to the familiar states 
$ u (r)
\propto
\sqrt{|E_{n}| \, r}  
\;
K_{i \Theta } (\sqrt{|E_{n}|} \, r)
$,
in terms of the Macdonald function~\cite{macdonald_function},
as shown in Refs.~\cite{renormalization_CQM,EFT_singular}.
In essence, a {\em properly physical quantum anomaly\/} 
is realized for strong coupling, where binding occurs 
[as in Eq.~(\ref{eq:energy-spectrum_CQM})] 
and the S-matrix~(\ref{eq:S-matrix_renormalized}) is 
manifestly scale dependent. Nevertheless, an important restriction is in order:
Eqs.~(\ref{eq:S-matrix_renormalized}) and (\ref{eq:energy-spectrum_CQM}) 
are limited in energy by the effective renormalization procedure,
thus guaranteeing the physical boundedness from below
of the effective Hamiltonian. 

The above prediction of the effective renormalization method 
can be compared with its counterpart for the self-adjoint extensions.
The latter can be obtained in the strong-coupling regime through a 
generalization of Eq.~(\ref{eq:S-matrix_SA}) for $s = i \Theta$; 
by inspection of the near-singularity behavior 
in Eq.~(\ref{eq:wave-function_SA-extension-basis-to-near-sing}), 
it follows that
Eq.~(\ref{eq:F-plus-minus}) still applies, with the modified functions
$\phi_{\pm} (s = i \Theta) = \pm i \pi \Theta /4 $.
As expected, the S-matrix~(\ref{eq:S-matrix_SA}) retains 
a $z$-parameter ambiguity but is otherwise similar to
Eq.~(\ref{eq:S-matrix_renormalized}) in its analytic structure.
Not surprisingly, the determination of its poles again involves the 
complex root required by inversion of the exponent $i \Theta$.
The ensuing symmetry-breaking strong-coupling energies
are structured with an infinite multiplicity factor
$e^{- 2 \pi n/\Theta  }$, which is the analogue of the phase factor 
in Eq.~(\ref{eq:S-matrix_poles}), but for an imaginary index.
Therefore, the energy spectrum of the self-adjoint extensions 
is formally identical to that of Eq.~(\ref{eq:energy-spectrum_CQM}),
where the real nature of the energies is guaranteed by the fact that
${\varrho}^{({\rm SA})} (z) $ 
is a pure phase in this coupling regime. 
Thus, at first sight in Fig.~\ref{fig:conformal_spectrum},
both methods appear to give the geometric conformal tower.
However, a critical difference should be highlighted: while an effective 
renormalization approach has a lower bound or base energy value by 
construction (as dictated by the ultraviolet physics), {\em the self-adjoint 
extensions generate a ``conformal tower'' devoid of lower bound\/}.
Thus, with reference to Fig.~\ref{fig:conformal_spectrum}, 
the complete tower extending to minus infinite energy
corresponds to the self-adjoint extensions,
while the renormalized spectrum further breaks the symmetry 
by terminating the sequence from below.
In short, from the conformal SO(2,1) symmetry viewpoint:
(i) 
the full-fledged {\em residual discrete symmetry\/}
is realized in the self-adjoint extensions;
(ii)
in the physical spectrum with energies~(\ref{eq:energy-spectrum_CQM}),
the residual discrete symmetry is terminated from below,
within an effective-field theory interpretation.

In short, while a spectrum not bounded from below is mathematically harmless,
such solution is physically untenable as it signals 
an unavoidable instability of the quantum system. This property 
ultimately persists because the self-adjoint extensions maintain 
the unadulterated form of the singularity---this is seen to be one aspect of
 the behavior known as ``the fall to the center''~\cite{landau:77}.
By contrast, the effective renormalization discussed above restores 
the physically mandatory boundedness, in another example of a phenomenon called
``quantum mechanical resuscitation''~\cite{stevenson:84}.

\section{\large\bf C\lowercase{onclusions and Outlook}}
\label{sec:conclusions}

 In this paper we have shown that the often-assumed equivalence between 
self-adjoint extensions and renormalization for singular potentials
fails to be valid for long-range interactions. Our conclusions 
broadly show that a physically-based analysis of singular potentials
resolves the physical indeterminacy of self-adjoint extensions
within the modern effective renormalization paradigm.

For the long-range conformal interaction, we have found major 
differences between the two methods. First, in the medium-weak 
coupling regime, the self-adjoint extensions typically include 
a bound state. By contrast, a physical regularization performs
a {\em medium-weak anomaly suppression\/} compatible with the
conformal symmetry, fully resolving the physical indeterminacy of 
self-adjoint extensions. Second, in the strong-coupling regime,
despite the formal similarity of the respective conformal towers
of bound states, the self-adjoint-extended Hamiltonians fail
to be bounded from below, unlike their renormalized counterparts.

The physical consequences of our analysis are noteworthy even from the
phenomenological viewpoint. First, a hypothetical realization of the 
medium-weak bound state would effectively imply
the loss of the critical nature of $\lambda^{(*)}$.
Thus, the observed sharp criticality, revealed by the absence of 
the medium-weak state, is a confirmation of the predicted 
anomaly suppression:
(i)
dipole-bound anions~\cite{molecular_dipole_anomaly} 
have an experimentally verified critical dipole moment 
$p^{(*)} \neq 0 $;
(ii)
in QED$_{3}$ with $N_{f}$ Dirac-fermion flavors~\cite{gusynin1},
the critical fermion number
$N^{(*)} = 128/( 3  \pi^{2} ) $ 
is given by conformal quantum mechanics and
also expected from alternative theoretical frameworks---moreover,
this is a feature of dynamical chiral symmetry breaking in gauge theories,
including applications in extra dimensions~\cite{gusynin2}.
Second, in the strong-coupling sector,
the {\em bounded tower of conformal states\/}~(\ref{eq:energy-spectrum_CQM})
with a sharp critical coupling $ \lambda^{(*)}  $ 
is the essence of the Efimov effect for three-body 
contact interactions~\cite{efimov_effect,Macek:06} 
and of the realizations discussed above.
Third, the medium-weak bound state has been invoked in 
black hole thermodynamics~\cite{near_horizon} through the 
static limit leading to a critical coupling
for the near-horizon physics; however, given the physical indeterminacy
of the self-adjoint extension method, one should not 
ascribe any fundamental meaning to these calculations---as it turns out, 
the effective near-horizon quantum-mechanical 
equations~\cite{BH_thermo_CQM,semiclassical_BH}
involve the strong-coupling sector.
Finally, the medium-weak regime may be
relevant for the question of cosmic censorship and the
geometric singularity of black holes, due to
the universal conformal behavior $\sim 1/x^{2}$ 
of the near-singularity field equations
with respect to the tortoise coordinate~\cite{Papdopoulos}; however, 
our paper suggests that caution must be exercised
in the use of self-adjoint extensions for the near-singularity physics.

Incidentally, many of the ideas of the previous sections---using 
a physical regularization with an ultraviolet core---admit a straightforward
generalization to {\em long-range strictly singular interactions\/}
$V({\bf r}) = - \lambda
\,
 r^{-\gamma}$,
with $\gamma > 2$ and $\lambda > 0$ in Eq.~(\ref{eq:singular-QM_radial}).
However, this generalized model is not scale invariant
and the dimensions of $\lambda$ play a central role;
with the standard arguments, considering this as a 
$(0+1)$-dimensional {\em effective\/} field theory~\cite{EFT},
one expects that some of the advantages of the 
renormalizable conformal interaction may be lost when $\gamma > 2$.
Thus, the competition between the kinetic and potential terms manifests in
a different manner: as the physics is ultimately modified 
in the presence of an ultraviolet scale $L_{UV}$ or cutoff $a$, 
a comparison of the competing terms yields the condition 
$ \lambda^{1/(\gamma -2)}  \gg L_{UV} \sim a $.
In addition, the semiclassical approximation is asymptotically exact 
with respect to the near-singularity behavior~\cite{semiclassical_BH};
consequently, the solutions follow from the usual WKB outgoing/ingoing 
near-singularity waves $u_{\pm}(r)$,
combined with a matching procedure with a regularizing core
for the removal of divergences---this is known to involve a 
limit-cycle behavior~\cite{beane:01}. Then, 
$\Omega \propto \exp \! 
\left[
-  4 \, i \, \lambda^{1/\gamma}
\, \left( k^{ 1 - 2/\gamma } -\mu^{ 1 - 2/\gamma } \right)/
\left(\gamma - 2 \right)
\right] 
$,
with $E_{0}= - \mu^{2} $ being an energy scale;
in turn, the S-matrix~(\ref{eq:S-matrix_singular})
reproduces the relevant physics, with
its poles corresponding to the leading-order renormalized energies 
$
 E_{ n }/ E_{0}  
= \left[ 1 - 
\left(  \gamma - 2 \right) \, \pi n  /
\left( 2 \, \lambda^{1/\gamma} 
\left| E_{0} \right|^{ ( \gamma - 2)/2 \gamma } \right)
\right]^{2 \gamma/\left( \gamma -2 \right) }
$
and reduces to the conformal counterparts in the limit $\gamma \rightarrow 2$,
with $\sqrt{\lambda} \rightarrow \Theta $
(the Langer replacement~\cite{semiclassical_BH}).
Therefore, for the strictly singular long-range interactions,
one concludes that:
(i) the critical coupling is zero;
(ii) there is no analog of the medium-weak coupling;
(iii) the strong-coupling regime is similar to the conformal
one, but without scale symmetry, with an ultraviolet cutoff suppressing
the unphysical behavior of a spectrum originally not bounded from below.
Relevant examples in molecular systems~\cite{vogt_wannier} ($\gamma = 4 $)
and for the nucleon-nucleon interaction~\cite{beane:01} ($\gamma =3 $)
are well known. The case with $\gamma =4$ can also be found in
extremal black holes and D-brane dynamics~\cite{park_D-brane_IQP}
and will be discussed elsewhere.

\acknowledgments{This work was supported by: 
the National Science Foundation under Grants No.\ 0308300 
and 0602340 (H.E.C.) and under Grants No.\ 0308435 and 0602301 (C.R.O.);
the 2006-2007 Fulbright Scholar Program (Award 6496)
and the University of San Francisco Faculty Development Fund
(H.E.C.); and CONICET and ANPCyT, Argentina (L.N.E., H.F., and C.A.G.C.).
We thank Jos\'{e} Garc\'{\i}a Esteve for his interesting comments.
}

\newpage

\bigskip


\end{document}